\documentclass[twocolumn,floatfix,superscriptaddress,aps,prl,noeprint]{revtex4-2}

\usepackage{dsfont} % Typesetting the mathematical symbols for the natural numbers (N), whole numbers (Z), rational numbers (Q), real numbers (R) and complex numbers (C).
\usepackage{bm} % \bm to makfe bold fonts in math mode

\usepackage{bbold} %For identity
\usepackage{xcolor}
\usepackage{physics}
\usepackage{graphicx,braket}
\usepackage{hyperref}
\hypersetup{colorlinks=true,linkcolor=blue,citecolor=blue}
\usepackage{amsmath,amssymb,amsfonts}
\usepackage{wrapfig}
\usepackage{multirow}
\usepackage{siunitx}

\def\ket#1{\mathinner{|{#1}\rangle}} 

\usepackage{cleveref}% load last
\crefname{equation}{Eqs.}{Eqs.}
\Crefname{equation}{Equation}{Equations}% For beginning \Cref
\crefrangelabelformat{equation}{(#3#1#4--#5#2#6)}
\crefmultiformat{equation}{Eqs. (#2#1#3}{, #2#1#3)}{#2#1#3}{#2#1#3}
\Crefmultiformat{equation}{Equations (#2#1#3}{, #2#1#3)}{#2#1#3}{#2#1#3}
\usepackage{ amssymb }

\DeclareMathOperator*{\argmaxA}{arg\,max} 

\begin{document}

\flushbottom
\title{Parameter estimation by learning quantum correlations in continuous photon-counting data using neural networks}

\author{Enrico Rinaldi}
\affiliation{Quantinuum K.K., Otemachi Financial City Grand Cube 3F, 1-9-2 Otemachi, Chiyoda-ku, Tokyo, Japan}
\affiliation{Interdisciplinary Theoretical and Mathematical Sciences (iTHEMS) Program, RIKEN, Wakoshi, Saitama 351-0198, Japan}
\affiliation{Center for Quantum Computing, RIKEN, Wakoshi, Saitama 351-0198, Japan}
\affiliation{Theoretical Quantum Physics Laboratory, Cluster for Pioneering Research, RIKEN, Wakoshi, Saitama 351-0198, Japan}

\author{Manuel González Lastre}
\affiliation{Departamento de Física Teórica de la Materia Condensada and Condensed
Matter Physics Center (IFIMAC), Universidad Autónoma de Madrid, 28049 Madrid,
Spain}
\affiliation{Instituto Nicolás Cabrera, Universidad Autónoma de Madrid, 28049 Madrid,
Spain}
		
\author{Sergio García Herreros}
\affiliation{Departamento de Física Teórica de la Materia Condensada and Condensed
Matter Physics Center (IFIMAC), Universidad Autónoma de Madrid, 28049 Madrid,
Spain}
\affiliation{Instituto Nicolás Cabrera, Universidad Autónoma de Madrid, 28049 Madrid,
Spain}			

\author{Shahnawaz Ahmed}
\affiliation{Department of Microtechnology and Nanoscience, Chalmers University of Technology, 412 96 Gothenburg, Sweden}

\author{Maryam Khanahmadi}
\affiliation{Department of Microtechnology and Nanoscience, Chalmers University of Technology, 412 96 Gothenburg, Sweden}

\author{Franco Nori}
\affiliation{Center for Quantum Computing, RIKEN, Wakoshi, Saitama 351-0198, Japan}
\affiliation{Theoretical Quantum Physics Laboratory, Cluster for Pioneering Research, RIKEN, Wakoshi, Saitama 351-0198, Japan}
\affiliation{Physics Department, University of Michigan, Ann Arbor, MI 48109-1040, USA}

\author{Carlos S\'anchez Mu\~noz}
\email{carlossmwolff@gmail.com}
\affiliation{Departamento de Física Teórica de la Materia Condensada and Condensed
Matter Physics Center (IFIMAC), Universidad Autónoma de Madrid, 28049 Madrid,
Spain}
\affiliation{Instituto Nicolás Cabrera, Universidad Autónoma de Madrid, 28049 Madrid,
Spain}

\begin{abstract}
We present an inference method utilizing artificial neural networks for parameter estimation of a quantum probe monitored through a single continuous measurement. Unlike existing approaches focusing on the diffusive signals generated by continuous weak measurements, our method harnesses quantum correlations in discrete photon-counting data characterized by quantum jumps.
We benchmark the precision of this method against Bayesian inference, which is optimal in the sense of information retrieval. By using numerical experiments on a two-level quantum system, we demonstrate that our approach can achieve a similar optimal performance as Bayesian inference, while drastically reducing computational costs. Additionally, the method exhibits robustness against the presence of imperfections in both measurement and training data. This approach offers a promising and computationally efficient tool for quantum parameter estimation with photon-counting data, relevant for applications such as quantum sensing or quantum imaging, as well as robust calibration tasks in laboratory-based settings.
\end{abstract}
\date{\today} \maketitle
\section{Introduction}
Quantum parameter estimation refers to the fundamental problem of inferring the value of unknown physical quantities within a quantum system~\cite{wiseman_book10a}. It has important practical applications in device characterization~\cite{genois2021, carrasco2021}, quantum feedback and control~\cite{zhang2017,wiseman1993a,ashhab2010,sayrin2011,yonezawa2012,cui2013} and quantum communications~\cite{leverrier2015,liang2022}, and most importantly, it is at the core of the field of quantum metrology~\cite{Giovannetti2006, giovannetti2011,degen2017, pirandola2018,pezze2018, ma2011, liu2016b, xu2022, lee2023}, whereby quantum correlations are exploited to estimate physical quantities with high precision.

The most conventional approach to quantum parameter estimation is based on the following steps: i) prepare a quantum system in a given state; ii) allow it to evolve through a unitary transformation that encodes the unknown parameters; iii) perform a measurement on the system; iv) estimate the parameter from the measurement results; v) repeat the process to gather statistical information and increase the precision of the estimation~\cite{degen2017,pezze2018}. Beyond this standard prepare-evolve-measure approach, there are relevant situations where, instead, one performs a single-shot continuous quantum measurement on a quantum probe~\cite{wiseman_book10a}. In this case, one is interested in obtaining an estimation that will be continuously updated  as data is gathered over the time evolution of a single quantum trajectory. Protocols of quantum parameter estimation from continuous measurements have been developed and studied over the years
~\cite{mabuchi1996,Gambetta2001,verstraete2001,chase2009,guta2011,ralph2011,Gambetta2001,Gammelmark2014, Gammelmark2013,Kiilerich2014,kiilerich2015,kiilerich2016,cortez2017,ralph2017,albarelli2018}; such parameter estimation techniques are particularly relevant for applications such as atomic magnetometry~\cite{kominis2003,geremia2003}, quantum metrology with open quantum systems~\cite{Gammelmark2014,macieszczak2016,ilias2022} or estimation of classical time-dependent signals~\cite{tsang2011}. 
Previous works have established computational approaches to calculate the probability density $P(\theta|D)$~\cite{Gambetta2001,Gammelmark2013,Kiilerich2014,kiilerich2015,kiilerich2016} of unknown $n$-dimensional parameters $\theta = [\theta_1,\theta_2,\ldots,\theta_n]$ conditioned on the continuously recorded data, $D$, which could correspond, for instance, to photon-counting records~\cite{Kiilerich2014,kiilerich2015}, continuous weak measurements~\cite{smith2006,ashhab2009,gong2018}, or homodyne measurements~\cite{kiilerich2016}.  These calculations rely on the application of Bayes' theorem on the likelihood functions $P(D|\theta)$, obtained as the trace of un-normalized conditional density matrices with the evolution conditioned on the measured data record~\cite{wiseman1993}.

This Bayesian approach provides, in principle, the most accurate description of the knowledge that can be gained about $\theta$ given the measurement data $D$. However, the application of this technique faces several difficulties. Firstly, it requires exact modelling of the open quantum system and any source of error, which can become challenging for noisy, complex quantum systems~\cite{chen2022}. Secondly, the method can be computationally demanding. For instance, when tackling the problem of multi-parameter estimation, where one wishes to estimate several parameters simultaneously, Bayesian inference calls for the use of stochastic methods to sample the posterior probability distribution~\cite{Gammelmark2013}.
Generally, the Bayesian parameter estimation method will be computationally time-consuming, even in simple systems. This hampers the prospects for real-time estimation and for the integration of the inference process in the actual measurement device taking the data, which would be desirable for reduced latency and energy consumption~\cite{verhelst2020,homrighausen2023}. 

A novel approach to this problem that has gained attention in recent years is the application of machine-learning methods in which a neural network (NN) is trained to aid in the inference process. Crucially, this method  allows to circumvent the requirement for detailed knowledge of the underlying physical system, as long as a reasonable amount of training data is available.
The use of NNs have been demonstrated for a variety of tasks in quantum metrology, including parameter estimation in the presence of unavoidable noise \cite{liu2019parameter,ban2021neural,nolan2021frequentist,xiao2022,ban2022},  calibration~\cite{cimini2019,cimini2021}, phase estimation in interferometry~\cite{cimini2019,cimini2021,cimini2023}, magnetometry~\cite{khanahmadi2021,ban2022}, control~\cite{xiao2022}, and tomography~\cite{ahmed2021,ahmed2021a}. 
In the context of continuously monitored systems, NNs have been employed for the reconstruction of quantum dynamics~\cite{flurin2020,genois2021} and parameter estimation
under continuous weak dispersive measurements~\cite{greplova2017,liu2019parameter,ban2021neural,khanahmadi2021}, which are characterized by a continuous, diffusive evolution of the quantum system~\cite{wiseman_book10a, bartolo2017}. 

Here, we assess the potential of deep learning to tackle the parameter estimation task from continuous measurements of the photon-counting type. These measurements are described by point processes that, rather than yielding a diffusive evolution of the system, are characterized by sudden quantum jumps upon the acquisition of a discrete signal, such as the detection of a single photon~\cite{wiseman_book10a}. This type of measurements, ubiquitous in the field of quantum optics and cavity-QED~\cite{zubizarretacasalengua2020, desantis2017, silva2016, rundquist2014, israel2017,lambert2010,hadfield2009,kimble77a}, provide insights into the particle-like nature of quantum systems and can produce strongly correlated signals that directly violate classical inequalities, as in the case of the renowned effect of photon antibunching~\cite{kimble77a}. Our work addresses the question of whether NNs can effectively process and exploit the quantum correlations present in this type of discrete signals.

\begin{figure}[t]
\includegraphics[width=0.48\textwidth]{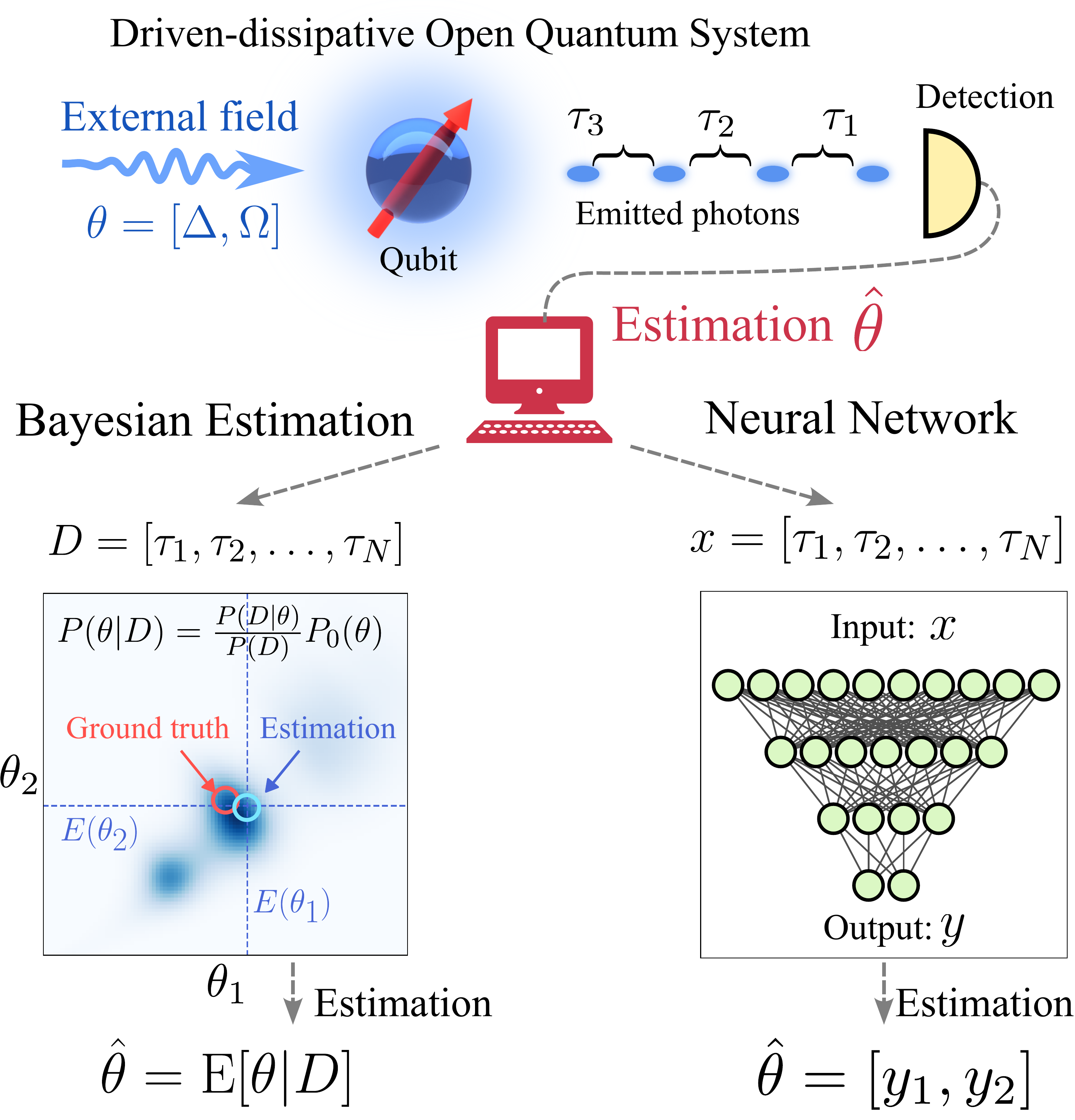}
\caption{\textbf{Quantum parameter estimation strategies in open quantum systems}. Parameters are encoded in the dynamics of an open quantum system: here, these are the frequency detuning between the qubit transition frequency and the drive frequency, $\Delta = \omega_q-\omega_L$, and the Rabi frequency $\Omega$ of the coherent drive, proportional to the dipole moment of the qubit and the amplitude of the electromagnetic field driving it. Photon-counting measurements of the quantum light radiated yield a record of the times of photodetection. The unknown parameters can be reconstructed through Bayesian parameter estimation (left). Here, we show the posterior probability distribution $P(\theta|D)$ for the data obtained from a single quantum trajectory, from which an estimator $\hat\theta$ is obtained by computing the mean value of the distribution. The ground truth value for the parameters is contained within the posterior distribution in a region of high probability. An alternative approach is based on the use of neural networks (right). The neural network learns the inverse process that maps observed data to the parameters $x \to \text{NN} \to \hat\theta$ through an initial training phase that employs data generated with known parameters. Once trained, inference for new data  is much faster using the neural network than repeating the Bayesian parameter estimation with the new data.}
\label{fig:fig1_setup}
\end{figure}

Our proposed NN architectures approach this task as a regression problem, taking the time delays between individual photodetection events as an input and providing an estimate $\hat\theta$ for the desired parameters of interest as the output, as sketched in Fig.~\ref{fig:fig1_setup}.
Our main finding is that the NNs can take advantage of the quantum correlations within the signal, providing higher precision than using an equivalent signal that lacks these correlations, e.g., the integrated intensity value. By computing the root-mean-squared error (RMSE) of the estimations, we find that, after training, the NN can reach the limit of precision established by Bayesian inference, while requiring dramatically fewer computational resources.
We also demonstrate that parameter estimation with NNs could be more resilient to noise than a simple Bayesian approach with no detailed knowledge of the underlying noise sources incorporated into the model, since it is possible for NNs to indirectly learn the noise model during the training phase. 
In contrast to a full Bayesian posterior distribution, the architectures studied in this work provide a single point estimate of the parameters, without including a confidence interval. Nevertheless, we evaluate their efficiency as estimators by conducting statistical analyses of their performance across numerous data samples.

\begin{figure*}[t!]
\includegraphics[width=0.98\textwidth]{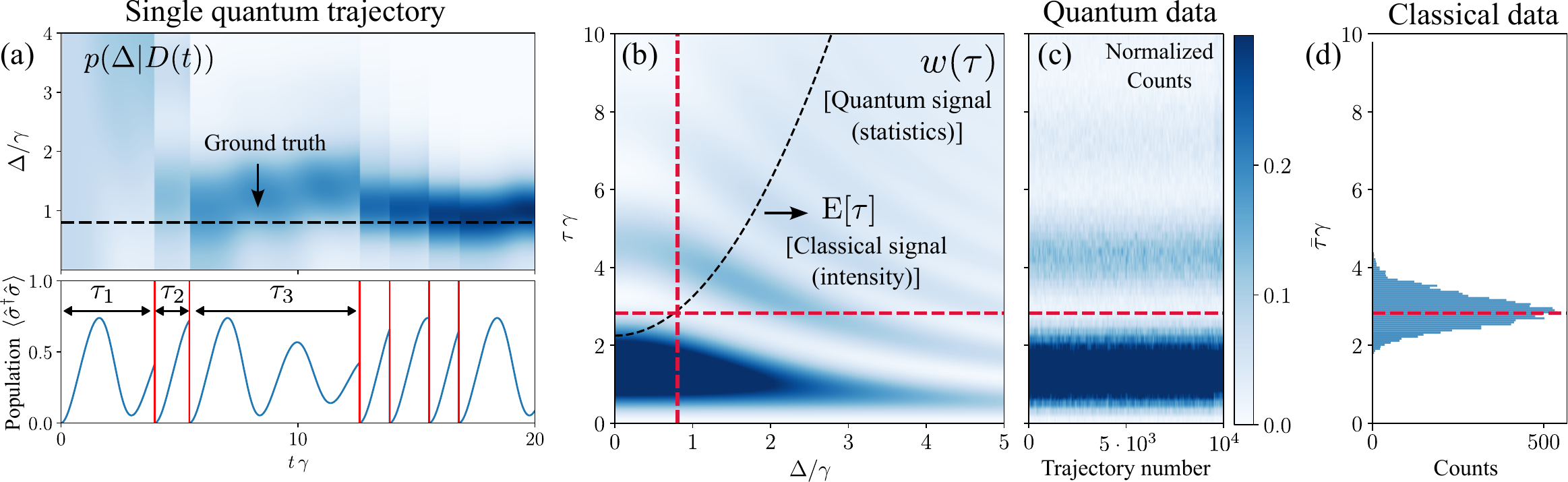}
\caption{\textbf{Classical and quantum data}. 
(a) Dynamics of an individual Monte Carlo quantum jump trajectory for $\Delta = 0.8\gamma$. Top panel shows the posterior probability of the parameter $\Delta$ for the photon-counting data $D(t)$ acquired up to time $t$ from a certain trajectory, starting from a flat prior. Bottom panel shows the corresponding evolution of the population of the qubit for the same trajectory, with red grid lines marking the times when a quantum jump occurs, leading to the collapse of the population to zero and subsequently restarting the system from the $\ket{0}$ state. The list of time delays between jumps constitutes the quantum data $D$ from which we infer the unknown parameters $\theta$. 
%At the times at which a quantum jump occur, the knowledge about the uknown $\theta = [\Delta]$  is significantly updated, as evidenced by the abrupt changes in the posterior. 
(b) Waiting-time distribution $w(\tau)$ of the measured time delays $\tau_i$ between photo-detection events during an individual trajectory, which constitute the quantum signal. The plot shows the distribution for several $\Delta$ parameters of the system. Clear patters emerge which are $\Delta$-dependent. 
A classical signal corresponding to the integrated intensity is defined as the mean of the measured delays, $\bar\tau = \sum_{i=1}^N \tau_i/N$. Since all $\tau_i$ are independent random variables coming from the same underlying distribution $w(\tau)$, $\mathrm{E}[\bar\tau]  = \mathrm{E}[\tau_i] = \mathrm{E}[\tau]$, shown as a black-dashed line. 
%The quantum signal has more $\Delta$-dependent features and therefore  carries more information than the classical one. 
The red vertical gridline marks the value $\Delta=0.8\gamma$ used as ground truth in panel (a). The horizontal grid line marks the corresponding expected value $\mathrm{E}[\tau](\Delta=0.8\gamma)$
(c) Heatmap of normalized histograms of $\tau$ obtained from $10^4$ trajectories with $N=48$ clicks each, simulated with $\Delta = 0.8\gamma$. The oscillatory patterns of the waiting-time distribution are clearly reproduced in the histograms, which represent the quantum data.
(d) Histogram of all the values $\bar\tau$ obtained for the $10^4$ trajectories used in panel (c). This represents the classical data, given by the mean of delays measured in a given trajectory. The resulting  Gaussian distribution is centered around $\mathrm{E}[\tau]$, marked by the dashed red gridline. We fix $\Omega = \gamma$ in all panels. All units are defined in terms of $\gamma$.}
\label{fig:fig2}
\end{figure*}

\section{Results}
\subsection{System and data generation}
The data we use to train the deep neural networks and perform parameter inference consists of simulated measurements of single photo-detection times from the light emitted by a driven two-level system (TLS).
This system represents a paradigmatic example of the types of quantum parameter estimation problems studied extensively in the literature~\cite{mabuchi1996,Gambetta2001,Gammelmark2014}. It is a prototypical model that can describe various quantum emitters, such as quantum dots~\cite{lodahl2015}, molecules~\cite{toninelli2021}, or color centres~\cite{sipahigil2016}.
We consider a TLS with states $\{|0\rangle, |1\rangle \}$,  lowering operator $\hat\sigma = |0\rangle \langle 1|$ and transition frequency $\omega_q$, continuously excited by a coherent drive with frequency $\omega_L$ and a Rabi frequency $\Omega$. This quantum probe is coupled to the environment, giving rise to spontaneous emission with a rate $\gamma$. In the frame rotating at the drive frequency, the dynamics of such a system is described by the master equation (we set $\hbar = 1$ herein):
\begin{equation}
    \partial_t \hat\rho = -i\left[\Delta \hat\sigma^\dagger \hat\sigma + \Omega(\hat\sigma + \hat\sigma^\dagger),\hat\rho\right] + \frac{\gamma}{2}D(\hat\sigma)[\hat\rho],
    \label{eq:master_equation}
\end{equation}
where  $\Delta \equiv \omega_q - \omega_L$ and $D(\hat x)[\hat\rho]\equiv 2 \hat x \hat\rho \hat x^\dagger - \left\{\hat x^\dagger \hat x,\hat\rho \right\}$ follows the definition of the Lindblad superoperator~\cite{breuer_book02a}. The interplay between continuous driving and losses eventually brings the system into a steady state. 

The estimation is performed using the data $D$ generated by the continuous photon-counting measurement of the radiation emitted by the system during a single run of the experiment, considering an initial state $|\psi(0)\rangle = |0\rangle$ and a total evolution time, $T$, much longer than the relaxation time towards the steady state. We simulate this process using the Monte Carlo method of quantum trajectories~\cite{plenio1998,wiseman_book10a,wiseman1993} implemented in the QuTiP library~\cite{johansson2012,johansson2013}. 
Under continuous photon-counting measurements, the system evolves stochastically under the following rules:
\begin{itemize}
    \item at every differential time step, $dt$, the system can experience a quantum jump $|\psi(t+dt)\rangle \propto \hat\sigma|\psi(t)\rangle$ with a probability $p(t) = dt\gamma\langle\hat\sigma^\dagger \hat \sigma \rangle(t)$, which corresponds to the detection of a photon emitted by the system;
    \item otherwise, the system follows a non-Hermitian evolution $|\psi(t+dt)\rangle \propto [\mathbf{I}-i(\hat H - i \hat\sigma^\dagger\hat\sigma/2)dt]|\psi(t)\rangle$. The wavefunction is normalized in each time step.
\end{itemize}
Each stochastic simulation of a trajectory corresponds to an individual realization of the experiment, generating a data vector by storing the specific set of times $(t_1,\ldots, t_N)$ when a photodetection leading to a quantum jump occurs during the course of a total evolution time $T$, with $N$ being the total number of jumps recorded. For convenience, we define our data as the set of time delays between these jumps,  $D = [\tau_1,\ldots,\tau_N]$,  where $\tau_i = t_i -t_{i-1}$,  $i=1,\ldots N$ and $t_0 = 0$. An important feature of this particular system is that each time interval $\tau_i$ is an independent random variable, given that the system collapses to the ground state $|0\rangle$ after each emission~\cite{Kiilerich2014}.

Each trajectory is simulated for an evolution time $T$ chosen to generate a data record $D$ of precisely $N=48$ jumps. This number is chosen to meet a compromise; it is sufficiently large to ensure that the system is in a steady state, while also allowing for inference with a limited number of photons in a relatively short timescale spanning just a few tens of emitter lifetimes, which is typically within the range of nanoseconds for solid-state quantum emitters~\cite{desantis2017}. 
It is important to note that fixing $N$ requires that each trajectory will have a  different evolution time $T$~\cite{gneiting2021,gneiting2022}. 
More detailed information about the unravelling of the dynamics in quantum trajectories can be found in the Supplemental Material~\footnote{See Supplemental Material for further details on the theory of quantum jump trajectories, the choice of different estimators from posterior distributions, and the Fisher information and the CRB.}
%, which includes references~\cite{wiseman_book10a,bartolo2017,chen2022,khanahmadi2021,plenio1998, paris2009, cramer1999, matson2006}.}.

After obtaining the data $D$ from a single trajectory or experiment, our goal is to estimate the unknown parameters $\theta$ that govern the dynamics of the system. In this study, we specifically focus on the parameters $\Delta$ and/or $\Omega$, considering that $\gamma$ is known. While this might be an idealized setting, it is instructive enough to benchmark the power of quantum correlations and neural networks for the task of parameter estimation.
We are interested in benchmarking the NNs against the ultimate precision limit set by the Bayesian inference process. This process yields a posterior probability distribution for the unknown parameters conditioned on the measured data $D$: following Bayes' rule we have $P(\theta|D) = {P(D|\theta)}/{\int d\theta P(D|\theta)}$. A comprehensive explanation of Bayesian inference from continuous measurements is provided in Ref.~\cite{Gammelmark2014} and, within the context of this work, in Ref.~\cite{Note1}. 
An illustrative individual trajectory, generated through our simulation approach, is depicted in Fig.~\ref{fig:fig2}(a), which shows the evolution of the posterior probability distribution for a single parameter, namely $\theta = [\Delta]$, in the top panel, and the TLS population in the bottom panel. Red grid lines mark the times when a quantum jump takes place. 
One can see how the regions of high probability in the posterior distribution concentrate more around the ground truth 
as time increases and more data is collected.

Once the posterior distribution is obtained, we build an estimator by taking the mean, $\hat\theta = E[\theta|D] = \int d\theta P(\theta|D)\theta$. While reducing the full posterior to a single estimation $\hat\theta$ implies a loss of information and involves a subtle choice of the best estimator, this step allows us to straightforwardly compare the result with the estimation provided by the NNs.
Here we also note that we use a flat (uniform) prior on the parameters of interest, while in general the Bayesian framework allows us to introduce our knowledge of the system using arbitrary priors.

A central question of our study is whether NNs can leverage the quantum correlations in the data. The emission from a TLS is strongly correlated; the fact that the emitter can only store one excitation at a time gives rise to the well-known non-classical phenomenon of antibunching~\cite{kimble77a}, i.e., the suppressed probability of detecting two photons simultaneously. These quantum correlations can be evidenced through several observables, such as the waiting-time distribution $w(\tau)$ (the probability of two consecutive emissions to be separated by a time delay $\tau$), which presents a vanishing value at zero delay, $w(\tau=0)=0$, and an oscillatory behaviour that depends strongly on the system parameters. These features, shown in Fig.~\ref{fig:fig2}(b), are translated directly into the histograms of time delays observed in each individual trajectory, as we show in Fig.~\ref{fig:fig2}(c). We will refer to a signal containing these time delays as a quantum signal.

To assess the information contained in these correlations, we compare it to an equivalent classical signal where the only relevant value is the integrated intensity, i.e. the total number of counts detected per unit time,  $I = N/T$. We can interpret such a classical signal as measurements lacking the necessary time resolution to detect individual photons, providing an integrated intensity  without any access to the internal temporal correlations between photon emissions. In fact, we choose the inverse of the intensity as the classical signal, since it conveniently represents that we have reduced the list of measured delays, that constitute the quantum signal, by taking their average, $\bar\tau \equiv \sum_{i=1}^N\tau_i/N = I^{-1}$. 
Since $\bar\tau$ is a sum or independent random variables $\tau_i$ that all follow the probability distribution $w(\tau_i)$, the central limit theorem ensures that, when $N$ is large, $\bar\tau$ is itself a random variable that follows instead a normal distribution, with $\text{E}[\bar\tau]=\text{E}[\tau_i]$. This normal distribution stands in contrast to the more complicated probability distribution $w(\tau_i)$ followed by the individual random variables $\tau_i$. In the steady state, this distribution is centered around $\text{E}[\tau_i]=\text{E}[\tau]\equiv\int d\tau w(\tau)\tau = (\langle\hat\sigma^\dagger\hat\sigma\rangle_\mathrm{ss} \gamma)^{-1}$ [see Fig.~\ref{fig:fig2}(d)]. 

Access to $\text{E}[\tau]$ instead of $w(\tau)$ represents an obvious loss of information, where any contributions from quantum correlations is completely discarded.  
Thanks to this, the RMSE of the estimation made by using the classical signal serves as a reference for comparison when doing estimations with the full quantum signal: whenever the obtained RMSE is lower than that of the classical signal, we understand that the estimation method is taking some advantage from quantum correlations. The process of parameter estimation from this classical signal,  described in more detail the Methods section,  also follows a Bayesian approach in which the chosen estimator $\hat\theta$ is the mean of the posterior $P(\theta|\bar\tau)$.

\subsection{Training}
We considered different neural-network architectures with the common structure of inputs and outputs that we illustrate in Fig.~\ref{fig:fig1_setup}. The input consists of a vector $ x = D = [\tau_1, \tau_2, \ldots, \tau_N]$ of time delays between photodetection events, while the output  is a vector of size $n$ corresponding to the estimations of the $n$-dimensional estimation problem, $ y = [\hat\theta_1, \hat\theta_1, \ldots, \hat\theta_n]$. Here, we treat both one-dimensional estimation problems, with $\theta = [\Delta]$ in the 1D case, and $\theta = [\Delta, \Omega]$ in the 2D case. We fixed the size of the input vector to $N=48$, indicating that our networks perform estimations based on a single experiment run once 48 photons have been detected. 

Estimation via NNs requires a training stage akin to a calibration process (learning phase). In this work, the input data used for training, $x_\mathrm{train}$, consists of a list of time delays obtained from trajectories simulated in QuTiP, with the corresponding target data $y_\mathrm{train}$ being the ground truth parameters used to run the simulations. 
Our best results are achieved using two types of architectures:
\begin{enumerate}
    \item \emph{RNN}: A recurrent neural network (RNN) with two Long Short-Term Memory (LSTM) layers. 
    These networks can learn correlations in sequential data, rendering them particularly promising for exploiting temporal quantum correlations. It is worth noting that there are no such correlations between time delays in our present system, since the $\tau_i$ are independent due to the collapse following a quantum jump. Nonetheless, these networks still deliver a very good performance. 
    More information on these networks and their application in parameter estimation from continuous signals can be found in \cite{khanahmadi2021}.

    \item  \emph{Hist-Dense}: A fully connected neural network with a first custom layer which incorporates the physical knowledge of the absence of correlations between time delays by performing a histogram of the input data. The geometry of the bins of the histogram is defined by the number of bins and their range. We set these as non-trainable hyperparameters fixed at 700 total bins and a range given by $\tau_\mathrm{min}=0$ and $\tau_\mathrm{max} = 100/\gamma$. 
\end{enumerate}

\begin{figure}[t!]
\includegraphics[width=0.48\textwidth]{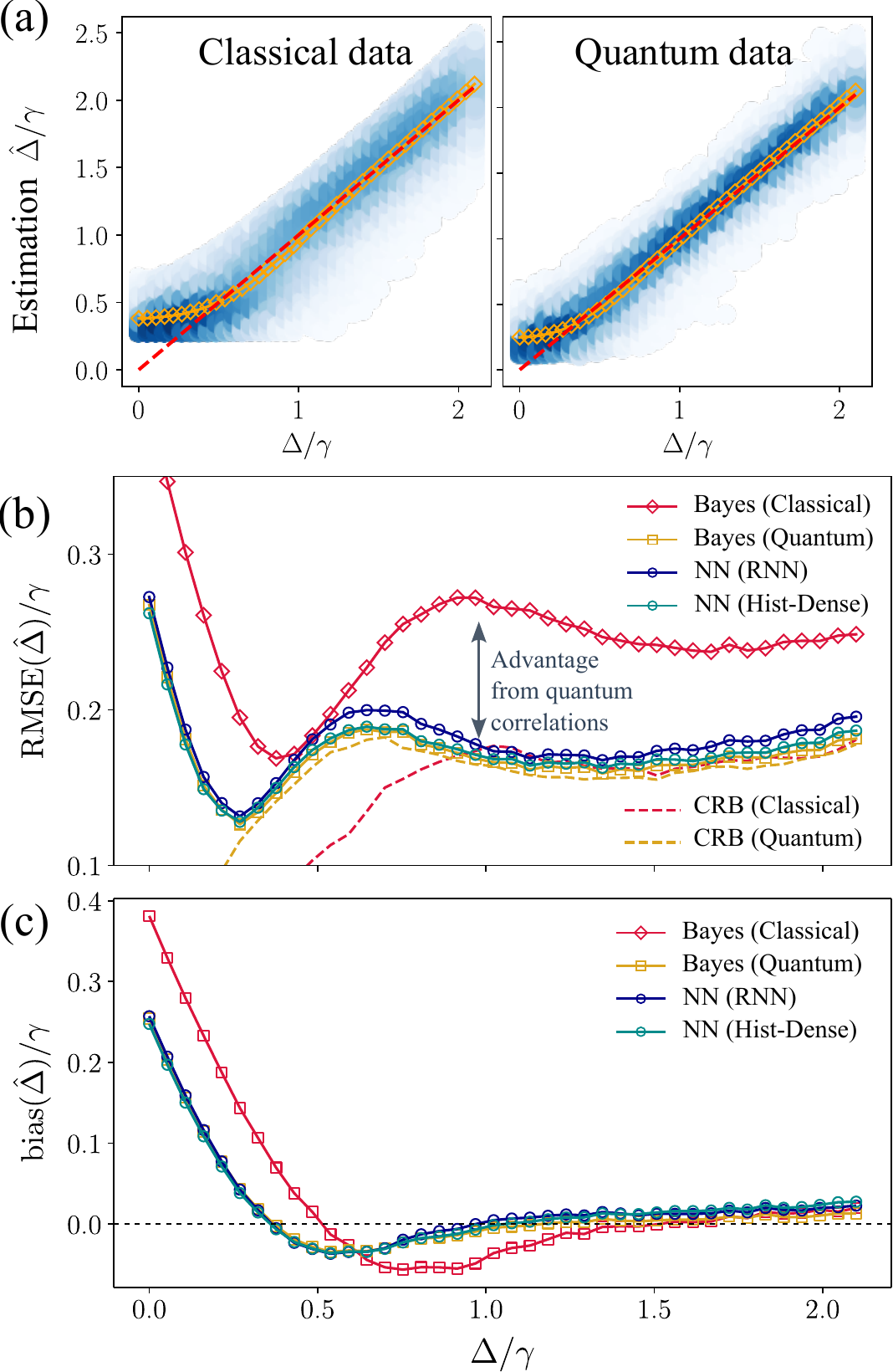}
\caption{ \textbf{Performance of different estimation strategies}. 
(a) Scatter plot of estimations  $\hat\Delta$ across a range of $\Delta$ values from classical (left) and quantum (right) versions of the photon-counting signal, computed for $10^4$ trajectories for each $\Delta$ value.  All the different estimators used for quantum data (Bayesian, NN) give a similar result, so only one is shown (NN-Hist). The blue shade of the points represents the underlying probability density function obtained through kernel density estimation. The dashed red line represents the ground truth. Orange points represent the mean of the the estimations. 
(b) RMSE of the estimators for on classical (red) and quantum (yellow, blue, green) versions of the signal. Estimations on quantum signals clearly outperform the classical estimation, signaling the useful information encoded in quantum correlations. The predictions made by NNs (blue, green) are on par with Bayesian inference (yellow), which is the optimal process in terms of information retrieval. Dashed lines mark the lower limit set by the biased Cram\'er-Rao bound for the classical and quantum signals.
(c) Bias of the predictions for the same cases as in (b).  }
\label{fig:MSE-1D}
\end{figure}

\begin{figure*}[t!]
\includegraphics[width=0.99\textwidth]{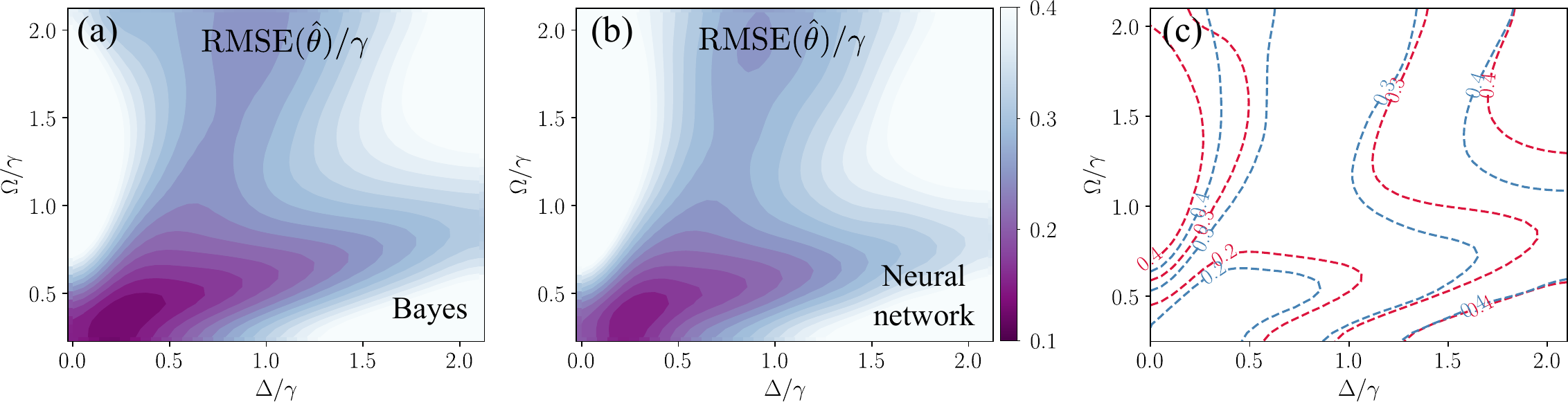}
\caption{\textbf{Multidimensional parameter estimation for $\theta = [\Delta, \Omega]$.} The qubit-drive detuning $\Delta$ and the Rabi frequency $\Omega$ are estimated simultaneously.  (a) RMSE of the estimations made through Bayesian inference. (b) RMSE of the estimations made by the NN. (c) Contours corresponding to Bayesian (red) and NN (blue) are overlaid for the sake of comparison. }
\label{fig:MSE-2D}
\end{figure*}

Further details on the networks and the training process can be found in the Methods section. In all the numerical calculations, the unit of time is set by $1/\gamma$ (which we assume to be a known parameter) by fixing $\gamma =1 $. After training, we benchmark the estimations made by the NNs against the estimation made through Bayesian inference and, in the single-parameter estimation case, also against the estimation done on the classical version of the signal. Details of this validation process are in the Methods section.

\subsection{Single-parameter estimation (1D)}
First, we focus on the simplest situation in which a single parameter is estimated: the detuning $\theta = [\Delta]$.
Figure~\ref{fig:MSE-1D}(a) depicts a scatter plot of the predictions made by each method for different values of $\Delta$. These plots show that the predictions made using the quantum signal---regardless of whether Bayesian inference or NNs are used---are more densely concentrated around the ground truth than the predictions obtained using the classical version of the signal. This already suggests that quantum correlations in the data can provide an advantage.

This advantage is clearly quantified in Fig.~\ref{fig:MSE-1D}(b), which compares the RMSE of the predictions obtained by each of the methods computed on a different set of validation trajectories (see Methods for details on the validation data). The estimation from classical data (red) always has a higher RMSE than estimation methods that have access to the quantum correlations in the data. Figure~\ref{fig:MSE-1D}(b) also shows that the NN architectures used in this work learn to extract information encoded in these correlations, with the resulting RMSE reaching values very close to the ultimate limit set by Bayesian inference. The fact that neural networks perform as well as the Bayesian approach by leveraging on quantum correlations in a completely model-agnostic manner is the main result of our work. 

We note, however, that none of the approaches studied here provided an estimator that is unbiased for all $\Delta$, as we show in Fig.~\ref{fig:MSE-1D}(c). There, we plot the bias of several estimators, defined as $\text{bias}(\hat\theta) \equiv \text{E}(\hat\theta - \theta)$. This bias is not problematic \emph{per se}, since it is known that biased estimators can be preferable to unbiased ones in some cases~\cite{jaynes03}. However, this bias needs to be taken into account at the time of comparing the obtained RMSE with the lower bound set by the Fisher information and the Cram\'er-Rao bound (CRB)~\cite{paris2009,cramer1999}. As we discuss in more detail in Ref.~\cite{Note1}, one needs to consider the \emph{biased} CRBs ~\cite{matson2006}. We calculated the Fisher information and the biased CRBs for the photon-counting measurements considered here~\cite{Gammelmark2013,Gammelmark2014,Kiilerich2014},
incorporating the calculated biases shown in Fig.~\ref{fig:MSE-1D}(c). The resulting lower limits for the RMSE are shown as dashed lines in Fig.~\ref{fig:MSE-1D}(b). We show that, for the quantum signal, both the Bayesian and NN estimators saturate the CRB for a wide range of values of $\Delta$.

\begin{figure*}[t!]
\includegraphics[width=0.98\textwidth]{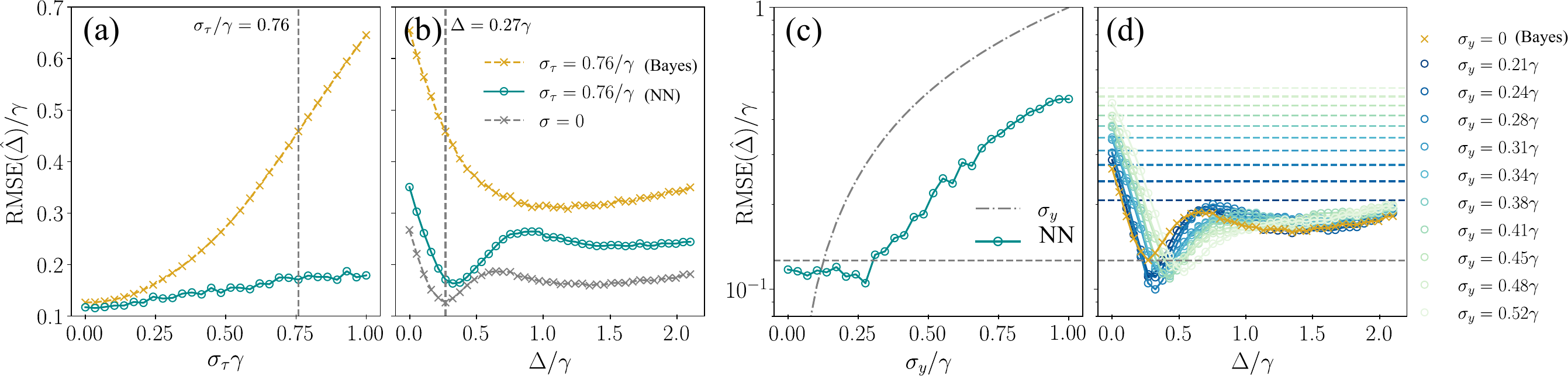}
\caption{\textbf{Parameter estimation with noisy data.} 
(a)
RMSE of NN predictor trained and evaluated with data containing jitter noise given by a Gaussian distribution with standard deviation $\sigma_\tau$ (green circles). This is compared to the RMSE of Bayesian inference lacking a description of the noise (yellow crosses). Results shown for a fixed $\Delta = 0.27\gamma$. 
(b) 
Same as in panel (a), shown versus $\Delta$ and with a fixed jitter noise $\sigma_\tau = 0.76\gamma$. RMSE of Bayesian predictor in the absence of noise shown for comparison (gray crosses). 
(c) 
RMSE of NN predictor trained with noisy target training data $y_\mathrm{train}$, with noise following a Gaussian distribution with standard deviation $\sigma_y$. We observe that the RMSE of the NN predictor (green circles) can remain smaller than the standard deviation of the training data $\sigma_y$ (gray dot-dashed line). The horizontal grid line marks the RMSE of Bayesian inference with noiseless data  for comparison . Results shown for a fixed $\Delta = 0.27\gamma$. 
(d) 
Sames as in panel (e), shown as a function of $\Delta$. Different colors represent different values of $\sigma_y$, which are also represented as horizontal grid lines, confirming that the RMSE of NN predictor remains smaller than $\sigma_y$ for all values of $\Delta$.}
\label{fig:noise}
\end{figure*}

\subsection{Multi-parameter estimation (2D)}

We also benchmarked our results for the 2D parameter-estimation case, where both $\Delta$ and $\Omega$ are unknown and estimated simultaneously, $\theta = [\Delta,\Omega]$. 

Our results for the 2D estimation case are shown in Fig.~\ref{fig:MSE-2D}. In panel (a), we depict the  RMSE of Bayesian estimation from the quantum signal in a region of the plane $[\Delta,\Omega]$. The resulting map shows a nontrivial structure that reflects the underlying physical complexity of the system and that is consistent with previous reports in the literature~\cite{Gammelmark2013,Kiilerich2014}.  The same physical complexity is captured by the NNs, as shown in Fig.~\ref{fig:MSE-2D}(b), which depicts the RMSE obtained by the estimation of a \emph{Hist-Dense} architecture and shows that the RMSE reflects the same underlying structure as the one obtained with Bayesian inference, yielding comparably small values overall. Both results can be more clearly compared in Fig.~\ref{fig:MSE-2D}(c), depicting overlaying contour lines of the two plots in panels (a) and (b).

\section{Discussion}

Our results clearly establish that NNs are able to exploit the information hidden in quantum correlations in order to perform parameter estimation.
Given their comparable precision to Bayesian inference, a careful evaluation of the benefits of opting for NNs as an alternative becomes important.

\subsection{Computation efficiency}
A first aspect in which NNs can bring an important advantage is the efficiency of the computation, which is particularly relevant for the case in which one wishes to achieve real-time estimation. For the 2D case of simultaneous estimation of two parameters, the posterior for a single trajectory for Bayesian estimation can be computed using the UltraNest package~\cite{2021JOSS....6.3001B}  (see Methods) in a typical wall time of 10 seconds on a MacBook Pro 2023 with an Apple M2 Max Chip. On the other hand, an estimation using the NNs built in this work, while being less informative (unlike the Bayesian posterior they provide a single estimation without a confidence interval) can be computed in a wall time of less than 1 ms, on the same device. 
The efficiency of the required calculation is such that trained NNs can be converted into compact and efficient TensorFlow Lite models, with a size of approximately 90kB. These models are suitable for deployment and execution on microcontrollers, which would minimize resource demands and enable data collection and estimation on the same device, leading to a significant reduction in latency ~\cite{verhelst2020,homrighausen2023}.
We can expect that, during data collection of photon-counting measurements, different chunks of measured time-delays are fed into the trained NN models for inference in real time.
Such a situation would not only allow for precise and robust real time inference, but could also inform about time-drifting parameters by looking at the temporal sequence of predictions.

\subsection{The challenge of noise}

Another important challenge faced by the Bayesian inference approach is the requirement of a precise modelling of the underlying physical system. This become particularly problematic in the presence of noise, which not only requires a perfect characterization and modelling of the different sources of imperfections, but also makes the subsequent calculation of posteriors much more involved. In contrast, the model-agnostic approach of using NNs allows for a robust prediction without the need of any modelling of the sources of noise, only by training the network directly on noisy data. 
In the particular case of photon-counting with single-photon detectors, numerous sources of noise can be attributed just to the detection process, such as the efficiency, bandwidth, dead time, dark counts or jitter noise~\cite{hadfield2009}. All these sources of noise have non-trivial consequences on the statistics of the photodetection events~\cite{lopezcarreno2022} and on the quantum trajectories associated to these imperfect measurements~\cite{warszawski2002}. Hence, the necessity of a perfect modelling and characterization of these processes seriously hampers the utilization of Bayesian inference. 

\subsection{Noise in \texorpdfstring{$x$}{x}: time jitter}

We illustrate the potential advantage brought in by the estimation with NNs by considering the particular case of time jitter~\cite{lopezcarreno2022}, which we describe by adding a noise term to the values of time delays $\tau$ used both in the training and estimation of the NNs for the 1D estimation case $\theta = [\Delta]$. We consider noise following a normal distribution with standard deviation $\sigma_\tau$, so that $x\rightarrow x + x_\mathrm{noise}(\sigma_\tau)$, with $x_\mathrm{noise}(\sigma_\tau) \sim \mathcal N (0,\sigma_\tau)$. 
In order to assess the resilience of the NN predictors to this type of noise, we defined a list of values of $\sigma_\tau$ in the range $\sigma_\tau=[0,\gamma^{-1}]$. For each value in this list, we sampled a population of $x_\mathrm{noise}$ and trained a model with the corresponding noisy data.

The resulting RMSE of the predictions returned by each of these models is computed using the same validation trajectories used in Fig.~\ref{fig:MSE-1D}, but containing an added Gaussian noise with the same $\sigma_\tau$ as the data used to train the model.  The resulting RMSEs are shown in Fig.~\ref{fig:noise}(a), compared to the RMSE of the prediction via an imperfect Bayesian inference process in which the noise is not explicitly modeled. We observe that, in the case of NNs, the results are more insensitive to the jitter noise and the RMSE of the NN predictor is always lower than that of the Bayesian predictor. This result remains true for all values of $\Delta$, as shown in Fig.~\ref{fig:noise}(b). We emphasize that this did not require any modelling of the imperfections, as would have been required for a correct process of Bayesian inference. The latter would presumably recover the values reported by the NN, but requiring a perfect characterization of the sources of noise, which may be challenging in some scenarios.

\subsection{Noise in $y_\mathrm{train}$: imperfect calibration}

Another relevant question is the effect of noise in the training target data $y_\mathrm{train}$. In particular, we consider the case in which the target data used for training differs from the underlying ground truth used to generate the quantum trajectories by a Gaussian noise term with standard deviation $\sigma_y$, so that $y_\mathrm{train}\rightarrow y_\mathrm{train} + y_\mathrm{noise}(\sigma_y)$, with $ y_\mathrm{noise} \sim \mathcal N(0,\sigma_y) $. This could reflect, for instance, the finite precision of a calibration device employed to experimentally measure the set of target values used to train the network.

The question we address is whether a NN will be able to perform parameter estimation with a lower RMSE than the standard deviation $\sigma_y$ of the noisy target data used for training. A positive answer to this question would indicate that the precision of the NN estimator can outperform that of the calibration method or device used to generate the training data. Figures~\ref{fig:noise}(c) and \ref{fig:noise}(d) illustrate that this scenario is indeed possible. Similarly to the process described for Figure~\ref{fig:noise}(a), we defined a series of values of $\sigma_y$ and, for each value, we sampled a population of $y_\mathrm{noise}$ and trained a model with noisy target data. 
The validation of the trained models is done using the same set of trajectories as in Fig.~\ref{fig:MSE-1D}.
Figure~\ref{fig:noise}(c) shows that, once the standard deviation of the noisy training target data $\sigma_y$ (dot-dashed line) becomes greater than the RMSE achievable in the noiseless case (horizontal grid line), the predictions made by the NNs trained with this noisy data remain robust and approximately equal to the ideal RMSE. As $\sigma_y$ increases, the RMSE of the NN predictor eventually increases as well, but it remains much smaller than $\sigma_y$. This observation holds for all values of $\Delta$, as shown in Fig.~\ref{fig:noise}(d). This result highlights the potential of the NNs to improve sub-optimal estimations used to feed the training process. 

\subsection{Conclusions and outlook}

 We have established the potential of NNs to perform parameter estimation from the data generated by continuous measurements of the photon-counting type. We have carefully benchmarked their precision against the ultimate limit set by Bayesian inference, showing that NN approaches can reach comparable performance in a much more computationally efficient way, while also being more robust against noise. 

These results pave the way for future research that can explore more complex quantum optical systems and measurement schemes. For example, future studies could investigate the radiation from cavity-QED systems with non-trivial quantum statistics~\cite{lambert2010,munoz2014,hamsen2017,zubizarretacasalengua2020} or several quantum emitters, as well as multiple simultaneous detection channels, such as those provided by arrays of single-photon avalanche diodes for generalized Hanbury-Brown and Twiss measurements~\cite{rundquist2014,desantis2017}, streak cameras~\cite{silva2016}, or single-photon fibre bundle cameras~\cite{israel2017,tenne2019}.

One aspect we leave for future work is the exploration of NN methods that incorporate mechanisms to do uncertainty quantification (UQ) on the predictions (compute prediction intervals)~\cite{nn-prediction-intervals}. For example, Bayesian Neural Networks~\cite{bayesian-dl-survey} train a set of weights distributions and the predictions from those networks are obtained by sampling the posterior of the weights after training. Other approaches to UQ include bootstrapping and ensemble methods which are easily implemented and only require extra training time, maintaining all the benefits of NN estimators that we have shown in our work. 

Another aspect deserving further exploration is leveraging domain knowledge and incorporating physical models into the learning process \cite{genois2021}. We note that the \emph{Hist-Dense} architecture used in this study could already be considered a physics-informed approach, taking advantage of a known feature about the model: the absence of information contained in the particular ordering of delays in the input data $x$, given that the source is a TLS that resets its state after every emission~\cite{Kiilerich2014}.

Moreover, it is worth noting that protocols of parameter estimation from photon-counting data, as the one described here, may find applications in fields such as fluorescence quantum microscopy~\cite{Delaubert2008,speirits2017,israel2017,tenne2019,he2023} or quantum imaging~\cite{albarelli2020, wolley2022}. In these areas, experimental quantum signals beyond the capabilities of classical simulations---due to the number of emitters and detectors involved---are already within reach, calling for efficient methods of analysis for the optimal image reconstruction~\cite{altmann2018}.

\section{Methods}

\subsection{Photodetection time data}
In practice, rather than working with the absolute times of detection, $t_i$, we find it more convenient to consider $D$ to be equivalently composed of a list of time delays  $\tau_i = t_i -t_{i-1}$,  with $i=1,\ldots N$ and $t_0 = 0$. 
For simplicity of calculations, we consider trajectories with different values of total evolution time $T$ but a fixed number of total detections $N$, and assume that the measurement is immediately terminated after the last detection, i.e., $t=t_N$. This means that $t-t_N$ will always be zero, becoming an irrelevant time interval that we do not need to include in our data, which will then only consist of $N$ values, $D= [\tau_1,\ldots \tau_N$]. This format  provides a set of data records with a fixed number of photodetection events or ``clicks'' that can be easily used as inputs for the NNs that we use in this work.

\subsection{Bayesian Inference}
\emph{Bayes' Rule.---} We consider that, at the beginning of our estimation protocol, our knowledge of the unknown set of parameters is characterized by a constant prior probability distribution $P_0(\theta)$ over an arbitrarily large but finite support (we define an initial support that is big enough not to affect the final estimation, and alternative prior distributions can also be used). Our goal is to update this knowledge upon the new evidence provided by some measurement data $D$, i.e., to compute the posterior probability $P(\theta|D)$. According to Bayes' rule, this is given by
\begin{equation}
P(\theta|D) = \frac{P(D|\theta)}{P(D)}P_0(\theta) = \frac{P(D|\theta)}{\int d\theta P(D|\theta)P_0(\theta)}P_0(\theta),
\label{eq:bayes}
\end{equation}
where the integral is defined over the domain of possible values of $\theta$, i.e., the support of $P_0(\theta)$. 
If we have no prior knowledge of $\theta$, as we assume here, $P_0(\theta)$ is constant over its support and can be taken out of the integral in the denominator in Eq.~\eqref{eq:bayes}, leaving us with a simple relation between $P(\theta|D)$ and $P(D|\theta)$:
\begin{equation}
\label{eq:bayes_simple}
P(\theta|D) = \frac{P(D|\theta)}{\int d\theta P(D|\theta)}.
\end{equation}
Thus, the required step to compute the posterior $P(\theta|D)$ is the calculation of the likelihood $P(D|\theta)$, which in general can be computed as the norm of a conditional density matrix evolving according to the list of quantum-jump times in data $D$, as explained in detail in Ref.~\cite{Note1}.

 \emph{Likelihood computation.---}
 The calculation of the likelihood is particularly simple for the dynamics of a TLS, since each time interval is independent of the others, given that the system is completely reset to the ground state $|0\rangle$ after each emission~\cite{Kiilerich2014}. Consequently, the probability for the set of data $D=[\tau_1,\ldots,\tau_N]$ is simply given by the product of the probabilities of each interval
\begin{equation}
P(D|\theta) = \prod_{i=1}^N w(\tau_i;\theta),
\label{eq:PDthetaTLS}
\end{equation}
where the waiting-time distribution $w(\tau;\theta)$ gives the probability of two successive clicks being separated by a time interval $\tau$ and we made explicit its dependence on the system parameters $\theta$. In our system, governed by the master equation \eqref{eq:master_equation}, the waiting-time distribution has the following analytical form,
\begin{multline}
    w(\tau; \theta)=\frac{8 \gamma  \Omega ^2 }{R}\exp[-\gamma  \tau /2]\\ \times \sum_{\eta=-1,1} \eta\cosh
   \left(\frac{\tau  \sqrt{\gamma ^2-4 \left(\Delta ^2+4 \Omega ^2\right)+\eta R}}{2
   \sqrt{2}}\right)
   \label{eq:waiting-time}
\end{multline}
 with $R=\sqrt{\left[\gamma ^2+4 \left(\Delta ^2+4 \Omega ^2\right)\right]^2-64 \gamma ^2
   \Omega ^2}$. 

\emph{Posterior computation.---} Once the method to compute the likelihood $P(D|\theta)$ is established, the only obstacle to using Bayes rule in Eq.~\eqref{eq:bayes} is the calculation of the normalization factor in the denominator.
This factor is called the evidence or the marginal likelihood, as it marginalizes the likelihood over all possible values of the parameters.
While this integral can be computed analytically for simple likelihood distributions, or numerically efficiently for a low-dimensional parameter space, many sampling techniques are typically used in Bayesian inference to bypass its direct computation.
Markov Chain Monte Carlo (MCMC)~\cite{speagle2020} is a stochastic sampling method that allows us to compute samples from the posterior $P(\theta|D)$ as draws from multiple Markov chains evolving in Monte Carlo time according to a stationary distribution proportional to the posterior.
Another method is Nested Sampling (NS)~\cite{buchner2023}, which is able to provide an estimate for the marginal likelihood itself by converting the original multi-dimensional integral over the parameter space to a one-dimensional integral over the inverse volume of the prior support.
NS methods have been shown to be effective also in the case of multi-modal posteriors (related to parameter degeneracy) and when the posterior distribution has heavy or light tails.

For the multi-dimensional case $n=2$, we compute the posterior probability distributions $P(\theta|D)$ using the UltraNest package~\cite{2021JOSS....6.3001B} in Python.
UltraNest contains several advanced algorithms to improve the efficiency and correctness of nested sampling, including bootstrapping uncertainties on the marginal likelihood and handling vectorized likelihood functions over parameter sets (together with additional MPI support).

%From the posterior distribution we can extract different estimators for the parameters which we compare to the machine learning estimators; for example, we can use the mean or the median of the posterior probability distribution as two estimators.

\subsection{Parameter estimation from classical signals}
We consider that the classical version of the measured quantum signal is given by the sample mean of the $N$ observed delays $\bar\tau = \sum_{i=1}^N \tau_i/N$. Being a sum of many independent random variables $\tau_i$, $\bar\tau$ is itself a random variable that, in the limit of large $N$, will follow a Gaussian distribution  $\bar\tau \sim \mathcal N(\mu,\sigma)$, where $\mu = \text{E}[\tau_i]$ and $\sigma^2 = \text{Var}[\tau_i]/N$. Since the independent random variables follow the waiting-time distribution given by Eq.~\eqref{eq:waiting-time}, $\tau_i \sim w(\tau;\theta)$, $\mu$ and $\sigma$ can be directly obtained:
\begin{subequations}
\begin{align}
  \mu(\theta) &= \left[\gamma \langle\hat\sigma^\dagger \hat\sigma\rangle_\text{ss}(\theta)\right]^{-1} = \frac{\gamma^2 + 4\Delta^2 + 8\Omega^2}{4\gamma\Omega^2},  \\
  \sigma(\theta)^2 &= \frac{(\gamma^2 + 4\Delta^2)^2 - 8(\gamma^2 - 12\Delta^2)\Omega^2 + 64\Omega^4}{N 16 \gamma^2 \Omega^4},
\end{align}
\end{subequations}
were we assume that the emission is recorded when the driven emitter has reached a steady state and we made explicit the dependence of these quantities on the unknown parameters $\theta$.
Within a Bayesian framework, this means that the likelihood of the classical data is
\begin{equation}
    P(\bar\tau|\theta) =\frac{1}{\sigma\sqrt{2\pi}} \exp\left[-\frac{1}{2}\left( \frac{\bar\tau - \mu}{\sigma}\right)^2\right],
    \label{eq:likelihood-classical}
\end{equation}
and the posterior is given by Eq.~\eqref{eq:bayes_simple}, assuming a constant prior. From here, we obtain a single estimation by taking the mean of the posterior, i.e., $\hat\theta = \text{E}[\theta|D] = \int d\theta \theta P(\theta|D)$. We make this choice in order to be consistent with the estimator used for Bayesian inference with quantum data, and because it shows overall lower RMSE and bias than the maximum-likelihood estimator, see Ref.~\cite{Note1}.
%

% As discussed in the main text, once a Bayesian posterior distribution $P(\theta|D)$ has been obtained, several estimator can be built. Since we assumed a constant prior for $\theta$, choosing the maximum-likelihood estimator (MLE) $\hat\theta_\mathrm{MLE} = \argmaxA_\theta P(D|\theta)$ or the maximum a posteriori estimator (MAP) $\hat\theta_\mathrm{MAP} = \argmaxA_\theta P(\theta|D)$ yields the same result. Given the likelihood in Eq.~\eqref{eq:likelihood-classical},  $\hat\theta_\mathrm{MLE}$ has an analytical solution too cumbersome to write here, but that approximately corresponds to the $\theta$ that solves of the equation $\mu(\theta) = \bar\tau$. For the 1D parameter estimation case, $\theta = [\Delta]$, such a solution is given by 
% %
% \begin{equation}
%     \hat \Delta_\mathrm{MLE} \approx \Re\left\{\frac{\sqrt{4 \gamma  T \Omega ^2-N \left(\gamma ^2+8 \Omega ^2\right)}}{2   \sqrt{N}}\right\}.
%     \label{eq:delta_estim_classical}
% \end{equation}
% %

\begin{table}[t!]
\centering
\begin{tabular}{l c c r}
\multicolumn{4}{l}{\emph{Recurrent Neural Network (RNN)}}\\
 \hline
\hline
Layer & Output shape & Activation & \# Parameters \\
\hline
LSTM  & 17 & ReLU & 1292 \\
LSTM & 17  & ReLU & 2380 \\
Dense & 1 & Linear & 18 \\
\hline
Trainable params. & 3,690 & &  \\
Epochs & 1200 & & \\ \\
\multicolumn{4}{l}{\emph{Hist-Dense}}\\ 
\hline
\hline
Layer & Output shape & Activation & \# Parameters \\
\hline
Histogram  & 700 & - & 0 \\
Dense & 100   & ReLU & 70100 \\
Dense & 50   & ReLU & 5050 \\
Dense & 30  & ReLU & 1530 \\
Dense &  1 & Linear & 31 \\
\hline
Trainable params. & 76,711 & &  \\
Epochs & 1200 & &
\\ \\
\multicolumn{4}{l}{\emph{Hist-Dense 2D}}\\ 
\hline
\hline
Layer & Output shape & Activation & \# Parameters \\
\hline
Histogram  & 700 & - & 0 \\
Dense & 100   & ReLU & 70100 \\
Dense & 50   & ReLU & 5050 \\
Dense & 30  & ReLU & 1530 \\
Dense & 20  & ReLU & 620 \\
Dense & 10  & ReLU & 210 \\
Dense &  2 & Linear & 22 \\
\hline
Trainable params. & 77,532 & &  \\
Epochs & 1200 & &
\end{tabular}
\caption{Summary of the NN architectures used for parameter estimation. All networks are trained with the Adam optimizer, with a learning rate $0.001$, using a batch size of 12800 and with a mean squared logarithmic error (MSLE) loss function. }
\label{table}
\end{table}

\subsection{Training and design of the neural networks}
For the 1D estimation case, the networks are trained with $4 \times 10^6$ trajectories in which the first 48 photodetection times are stored. From these, 80\% are taken as training data, and 20\% as validation data during the training process.
Each trajectory is simulated by fixing $\Omega = \gamma$, and choosing randomly a value of the detuning within the domain $\Delta \in [0,5\gamma]$, which should be large enough to cover the expected range of possible values of $\Delta$. 
We choose this domain to be the same as the support of the prior in the Bayesian inference, so that both methods rely on the same underlying assumptions about the probability distribution of $\Delta$. 
All the networks are trained using the Adam optimizer. A summary of these architectures---including number of trainable parameters and details about the batch size and number of epochs used in training---is provided in Table~\ref{table}. All the networks are defined and trained using the Tensorflow library~\cite{abadi2016}.

For the 2D parameter estimation, we train the network with a new set of $4\times 10^6$ trajectories  generated for random values of $\Delta$ and $\Omega$ in the range $\Delta\in [0., 3\gamma]$ and $\Omega \in [0.25\gamma,5\gamma]$ (lower values of $\Omega$ are not used in order to avoid longer simulations times associated with the lower pumping rate). 

\subsection{Validation process}
\subsubsection{1D case} In order to validate the performance of the NNs, we generate a set of validation trajectories for a uniform grid of 40 values of $\Delta$ (with $\Omega = \gamma$) in the range $\Delta \in [0,2.1\gamma]$. For each value of $\Delta$, we generated $10^4$ trajectories containing 48 photodetection events each. For each trajectory, we obtained an estimation of the parameter for each of the methods, taking the mean of the posterior as the estimation in the Bayesian case. We then computed the MSE from the $10^4$ different predictions with each of the methods for each value of $\Delta$.
\subsubsection{2D case} Validation is done for a grid of $(\Delta,\Omega)$ pairs built from a grid of 40 values of $\Delta$ in the range $\Delta\in [0., 2.1\gamma]$ and a grid of 40 values of $\Omega$ in the range $\Omega \in [0.25\gamma,2.1\gamma]$, spanning a square grid of 1600 points. For each of these points, we simulate $10^4$ validation trajectories with 48 clicks each.

\section{Data availability}
All the data necessary to reproduce the results in this work, including simulated quantum trajectories for training of the NNs, validation trajectories, trained models and Bayesian inference calculations in 2D via nested sampling are openly available on Zenodo~\cite{sanchezmunoz2023_dataset}.

\section{Code availability}
The codes used to generate the results in this work, including the implementation of the proposed NN architectures as TensorFlow models, are openly available on Zenodo and GitHub~\cite{sanchezmunoz2023_code}.

\bibliography{ParameterEstimation, references-enrico,books,REF-Maryam, QuantumImaging, software}

\section{Acknowledgments}
Authors thank Yexiong Zeng, Andrey Kardashin, Clemens Gneiting, Yanming Che, Fabrizio Minganti, Anton Frisk Kockum, Yuta Kikuchi, and Marcello Benedetti for critical reading and feedback.
E.~R. was supported by Nippon Telegraph and Telephone Corporation (NTT) Research during the early stages of this work.
S. A. and M. K. acknowledge support from the Knut and Alice Wallenberg Foundation through the Wallenberg Centre for Quantum Technology (WACQT). F. N. is
supported in part by Nippon Telegraph and Telephone
Corporation (NTT) Research, the Japan Science and
Technology Agency (JST) [via the Quantum Leap
Flagship Program (Q-LEAP) and the Moonshot R\&D
Grant No. JPMJMS2061], the Asian Office of Aerospace
Research and Development (AOARD) (via Grant No.
FA2386-20-1-4069), and the Foundational Questions
Institute Fund (FQXi) via Grant No. FQXiIAF19-06.
C.S.M. acknowledges that the project that gave rise to these results received the support of a fellowship from “la Caixa” Foundation (ID 100010434) and from the European Union’s Horizon 2020 research and innovation programme under the Marie Skłodowska-Curie Grant Agreement No.847648, with fellowship code LCF/BQ/PI20/11760026, and financial support from the MCINN project PID2021-126964OB-I00 (QENIGMA) and the Proyecto Sinérgico CAM 2020 Y2020/TCS- 6545 (NanoQuCo-CM).

\onecolumngrid
\appendix
%\FloatBarrier
\allowdisplaybreaks

\newpage

\newcommand{\beginsupplement}{
  \setcounter{table}{0}  
  \renewcommand{\thetable}{S\arabic{table}} 
  \setcounter{figure}{0} 
  \renewcommand{\thefigure}{S\arabic{figure}}
  \renewcommand{\theequation}{S\arabic{equation}}
  \setcounter{equation}{0}   

}

\beginsupplement
% \section{Theoretical Background}
% \label{appendixbackground}

\begin{center}
{\bf \large Supplemental Material}
\end{center}

\section{Unravelling dissipative dynamics into quantum trajectories}

To better understand the origin of this decomposition of the dynamics into trajectories experiencing quantum jumps, we may write the master equation in Eq.~\eqref{eq:master_equation} in superoperator form as $\partial_t \hat\rho = \mathcal L_{\theta} \hat \rho$, where $\mathcal L_{\theta}$ is the Liouvillian superoperator, and the subscript $\theta$ makes explicit its dependence on the unknown set of parameters we want to estimate, e.g., $\theta = [\Delta,\Omega]$, which are encoded into the state through the Liouvillian evolution.  We can separate the Liouvillian in two terms, $\mathcal L_{\theta}= \mathcal J + (\mathcal L - \mathcal J)$, where $\mathcal J$ represents action of a quantum jump onto the system~\cite{wiseman1993}:
\begin{equation}
\mathcal J \equiv \gamma  \hat\sigma \hat\rho \hat \sigma^\dagger.
\end{equation}
The unravelling into different trajectories becomes apparent by writing the density matrix at any time $T$ as a mixture of non-normalized conditional density matrices using a generalized Dyson expansion of the solution of the master equation,
\begin{equation}
\rho(T,\theta) = \sum_{N=0}^\infty \int_0^T dt_N \int_0^{t_N}dt_{N-1}\cdots \int_0^{t_2} dt_1
\\ \times\tilde{\rho} (t_1,\cdots,t_N;T;\theta),
\label{eq:me-dyson}
\end{equation}
each of these matrices being defined as
\begin{equation}
\label{eq:rho-conditional}
\tilde{\rho} (t_1,\cdots,t_N;T;\theta) = \mathcal S_{\theta}(T-t_N)\mathcal J \\
\times \mathcal S_{\theta}(t_N-t_{N-1})\cdots \mathcal{J} \mathcal{S}_{\theta}(t_1)\rho(0) \, ,
\end{equation}
where $\mathcal S_{\theta}(\tau) \equiv \exp[(\mathcal L_{\theta} - \mathcal J)\tau]$. This equation represents a conditional, un-normalized density matrix characterized by periods of evolution without a quantum jump generated by the operator $S_{\theta}$, and quantum jumps taking place at times $t_i$ through the operator $\mathcal J$.  
Equation~\eqref{eq:me-dyson} represents the unraveling of the evolution of $\rho$ in the time interval $[0,T]$ into all of possible individual realizations, i.e., the quantum trajectories. A trajectory is fully characterized by its duration, $T$, and the number of photons $N$ emitted during that time interval as well as their corresponding times of emission. In other words, the data array that contains all the photo-emission times, $ D = [t_1,\ldots, t_N]$ is enough to tag each individual quantum trajectory.
%Equation~\eqref{eq:me-dyson} reflects the fact the total density matrix is a mixture of all the possible realizations that can take place.
Equation~\eqref{eq:me-dyson} tells us that each of these realizations, tagged by the data array $D$, has a probability of occurring given by
\begin{equation}
P(D|\theta) = \text{Tr}\left[ \tilde \rho(t_1,\ldots,t_N;T,\theta \right].
\label{eq:P(D-theta)}
\end{equation}
This equation gives us the likelihood of the data. From this, we can  reconstruct the posterior probability distribution of the unknown parameters $\theta$ by direct application of Bayes' rule.

The unravelling of the master equation discussed previously establishes a connection to measurement theory and the concept of monitored realizations. When knowledge is acquired about the system, the state of the system is consequently updated, in accordance with the conditional density matrices presented in Eq.~\eqref{eq:rho-conditional}. It is important to note that there are multiple ways to unravel the dynamics, as the choice of the jump operator $\mathcal J$ can vary while still yielding the same master equation that we presented in \eqref{eq:master_equation}~\cite{bartolo2017}. Each different choice of the jump operator corresponds to a distinct unraveling and measurement strategy. Nevertheless, by averaging the ensemble of trajectories from each type of unraveling, we obtain the same density-matrix evolution, which does not assume that any knowledge has been retrieved.

Notable examples of different unravellings in quantum optics include those arising from photon-counting measurements, which exhibit abrupt jumps in trajectories, and homodyne measurements, which result in a diffusive evolution.
These two measurement strategies access the particle-like and wave-like nature of the electromagnetic field, respectively. Prior works utilizing neural networks in continuously monitored systems have primarily focused on the latter type of measurement, i.e. in continuous, diffusive signals~\cite{khanahmadi2021,chen2022}. In contrast, our work focuses on unravellings with abrupt jumps, corresponding to a ``click'' in the detector, producing discrete signals consisting of these precise photodetection times. 

A single trajectory describing the evolution of an individual run of the experiment in which the system is continuously monitored cannot be predicted deterministically, since its evolution is stochastic. This evolution preserves the purity of the state, allowing us to work in a framework that only uses wavefunctions. For a trajectory describing the specific case of continuous monitoring through photon-counting measurements, the wavefunction is governed by a nonlinear stochastic Schr\"{o}dinger equation~\cite{wiseman_book10a}
\begin{equation}
    d|\psi(t)\rangle = dN(t)\left(\frac{\hat\sigma}{\sqrt{\langle \hat\sigma^\dagger \hat\sigma\rangle(t)}} -\mathbf{I}\right)|\psi(t)\rangle \\
    + dt\left( \frac{\langle \hat\sigma^\dagger \hat\sigma \rangle (t)}{2} - \frac{\hat\sigma^\dagger \hat\sigma}{2} - i\hat H \right)|\psi(t)\rangle,
    \label{eq:SSE}
\end{equation}
where $dN(t)^2 = dN(t)$ and $E[dN(t)]=dt \langle \psi(t)|\hat\sigma^\dagger \hat\sigma |\psi(t)\rangle$. Notice that, following the discrete nature of photon-counting measurements,  $dN(t)$ is a discrete random variable that can be either 0 or 1, and that $E[dN(t)]$ corresponds to the probability of detecting a photon in the time window $(t,t+dt)$. Solving the previous equation by discretizing time leads to the usual recipe of Monte Carlo evolution described in the main text~\cite{plenio1998}.

\section{Choosing an estimator from the posterior}
One of the purposes of this work is to benchmark the precision of the single-point estimates produced by the proposed NN architectures against the estimation provided by Bayesian inference. Making an equivalent comparison between the two requires to build a single point estimator summarizing the information contained in the full Bayesian posterior distribution (which inevitably entails some loss of information). 
Moreover, we diminish the role of the prior in the Bayesian inference procedure by focusing on a uniform prior, since we do not equip the NN architectures with a mechanism similar to a prior distribution on parameters. It is well known that, when the data is scarce, the importance of a prior in a Bayesian protocol can be substantial.

Choosing the best estimator can be a subtle problem, since several single-point estimators can be built from the posterior. The most relevant ones are arguably the \emph{maximum a posteriori} (MAP) estimator, defined as
\begin{equation}
    \hat\theta_\text{MAP} =  \argmaxA_\theta P(\theta|D),
\end{equation}
and the \emph{mean estimator}, defined as the mean of the posterior
\begin{equation}
    \hat\theta_\text{MEAN} = \int d\theta \, \theta  P(\theta|D) .
\end{equation}
Notice that the MAP estimator is equivalent to the \emph{maximum-likelihood estimator} if there is no previous knowledge about the parameters, i.e., if the prior distribution $P_0(\theta)$ is flat.

\begin{figure*}[h]
\includegraphics[width=0.98\textwidth]{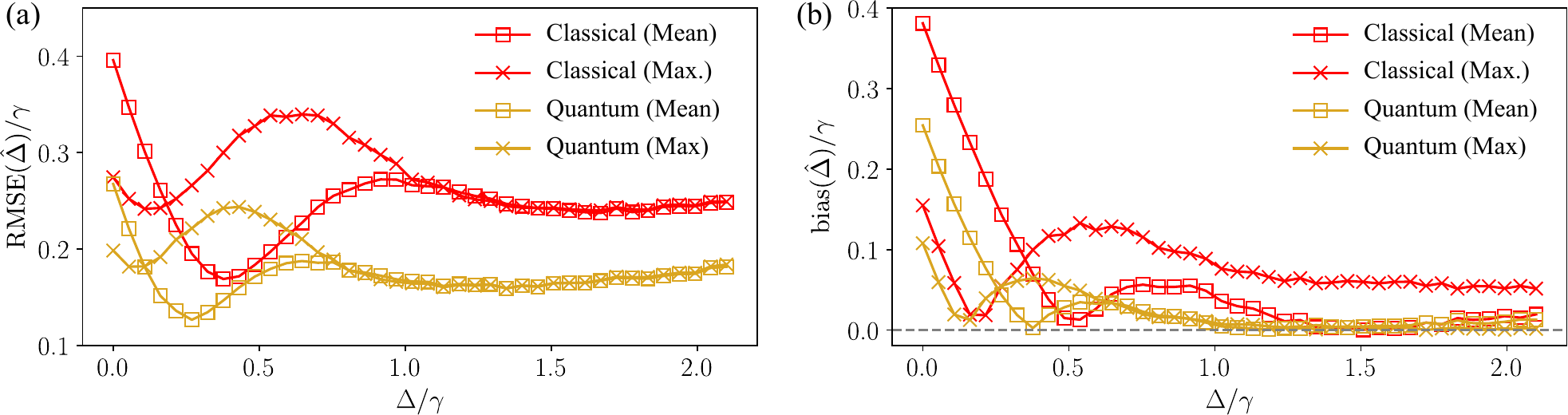}
\caption{\textbf{Performance of different estimators built from the Bayesian posterior distribution $P(\theta|D)$.} (a) RMSE of estimators obtained as the mean of the posterior distribution (squares) and as the maximum a posteriori probability (MAP) estimate (crosses). For most values of $\Delta$, the posterior mean estimate yields a lower RMSE. (b) Bias of the same estimators shown in (a).    }
\label{fig:figSM1}
\end{figure*}

In all the comparisons shown in this work, we have consistently selected the mean estimator $\hat\theta_\text{MEAN}$ as the one representing the performance of Bayesian inference, both for classical and quantum signals. We have done this without explicitly using the subscript when referring to $\hat\theta$, for the sake of notational simplicity. In this section, we justify our choice by showing in Fig.~\ref{fig:figSM1} the RMSE and bias of both MAP and mean estimators in the 1D estimation case, $\theta = [\Delta]$. These results show that, for most values of $\Delta$, mean estimators show a lower RMSE and bias than the MAP estimator, which justifies our choice of the mean estimator as the best representation of Bayesian inference. Notably, we find that the RMSE of the mean estimator coincides with the RMSE naturally reached by the NN estimators after their training.

\section{Fisher information}
Fisher information is ubiquitous as a tool to establish the ultimate limits of precision of parameter estimation approaches in quantum metrology~\cite{paris2009}. The Cram\'er-Rao bound (CRB) sets a limit for the variance of an \emph{unbiased} estimator given by~\cite{cramer1999}:
\begin{equation}
    \text{var}\hat\theta \geq \frac{1}{\eta F(\theta)}
    \label{eq:unbiased-CRB}
\end{equation}
where $\eta$ is the number of measurements and $F(\theta)$ is the Fisher information, 
\begin{equation}
    F(\theta ) = \int dx\, P(x|\theta)\left(\frac{\partial \ln P(x|\theta)}{\partial\theta}\right)^2,
    \label{eq:fisher}
\end{equation}
where we defined $x$ as the value of a measurement outcome and $P(x|\theta)$ as the probability to measure such a value given a specific value of the parameter $\theta$.

In the context of quantum measurements, the function $P(x|\theta)$ is determined by both the quantum state in which the measurement is performed and by the choice of measurement, established by a particular positive operator-valued measure (POVM), $\Pi$. For a given quantum state, the Fisher information is thus bounded by the \emph{quantum Fisher information} (QFI), $H$, which is the Fisher information corresponding to the optimal POVM:
\begin{equation}
    H = \max_\Pi F.
\end{equation}

The notions of Fisher information and quantum Fisher information have both been introduced and used in the context of continuous measurements in open quantum systems that we discuss in our work~\cite{Gammelmark2013, Gammelmark2014, Kiilerich2014}. In this case, the quantum state for which the quantum Fisher information is defined is the combined quantum state of system and environment~\cite{Gammelmark2013}, and the measurement strategy that saturates the Fisher information can, in general, be a very complicated measurement defined in this combined space. For a particular measurement strategy, $F$ can be calculated by computing $P(x|\theta)$ from Eq.~\eqref{eq:P(D-theta)}. The integral in Eq.~\eqref{eq:fisher} can then be approximated by sampling measurement outcomes, e.g., performing multiple simulations of quantum jump trajectories~\cite{Gammelmark2013}. 

\begin{figure*}[b]
\includegraphics[width=0.8\textwidth]{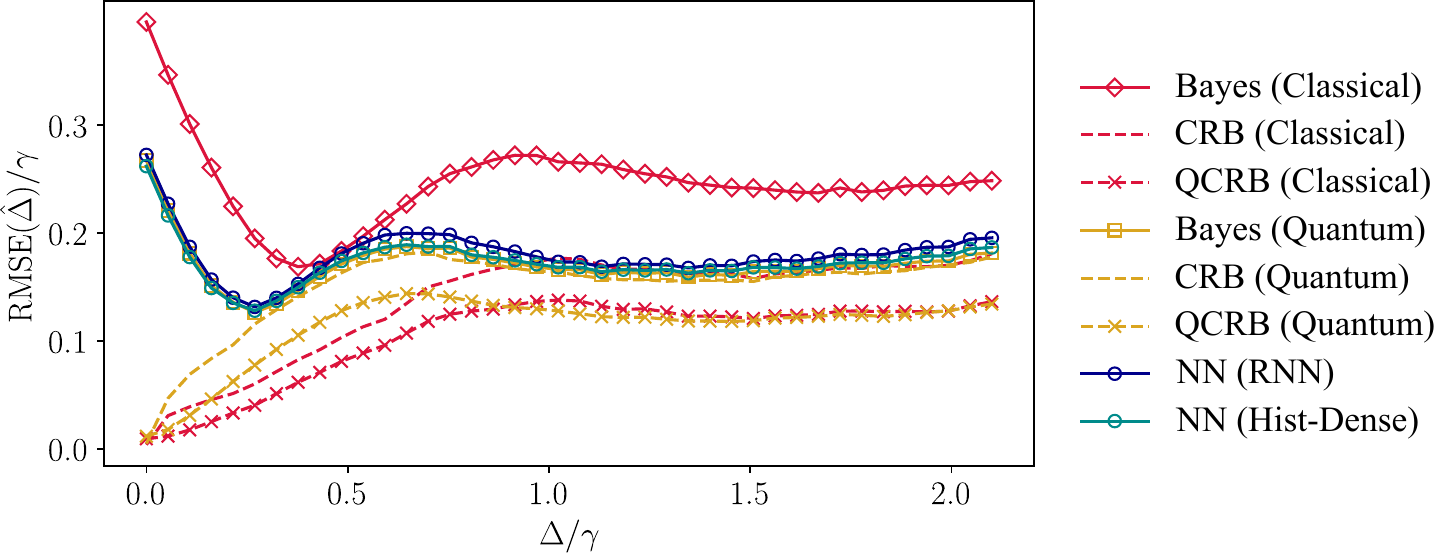}
\caption{\textbf{Limits set by the Cram\'er-Rao bound}. RMSE and bias of different estimators compared to the lower limit set by the biased CRB and QCRB, using the bias of each estimator.}
\label{fig:figSM2}
\end{figure*}

We have shown that all the estimators used in this work show some degrees of bias. In that case, the CRB defined by Eq.~\eqref{eq:unbiased-CRB} does not apply. Instead, we need to use a generalized \emph{biased} CRB, which is given by~\cite{matson2006}

\begin{equation}
    \text{var}\hat\theta \geq \left(1 + \frac{d \, \text{bias}(\hat\theta)}{d\theta} \right)^2\frac{1}{\eta F(\theta)}.
    \label{eq:biased-CRB}
\end{equation}
It is clear from this equation that estimators with a negative derivative of its bias will yield lower variance than that predicted by the unbiased CRB, as it is our case.
For reference, see the bias curves in Fig.~\ref{fig:MSE-1D}(c).

In order to assess how far our estimators are from the lower limit set by the CRB, we computed the Fisher information for photon counting, $F$, by calculating the likelihood functions $P(D|\theta)$ and their derivatives over samples of quantum trajectories, as discussed in more detail in Ref.~\cite{Gammelmark2013}. We also computed the quantum Fisher information, $H$, following the approach in Ref.~\cite{Gammelmark2014}, by which $H$ is given by
\begin{equation}
    H(\theta) = 4 T \partial_{\theta_1}\partial_{\theta_2} \text{Re}[\lambda_s (\theta_1, \theta_2)]_{\theta_1 = \theta_2 = \theta},
\end{equation}
where we set the time as the average duration of the trajectory $T =N/(\gamma \langle\hat\sigma^\dagger\hat\sigma\rangle_\mathrm{ss})$, and $\lambda_s(\theta_1,\theta_2)$ is the smallest eigenvalue of the generalized Liouvillian $\mathcal L _{\theta_1, \theta_2}$ which, in our case, is defined as 
\begin{equation}
   \mathcal L _{\theta_1, \theta_2}[\hat\rho] \equiv -i \left(\hat H(\theta_1)\hat\rho - \hat\rho \hat H(\theta_2) \right) + \frac{\gamma}{2}D(\hat\sigma)[\hat\rho],
\end{equation}
with $H(\theta) = \Delta \hat\sigma^\dagger \hat\sigma + \Omega(\hat\sigma + \hat\sigma^\dagger)$ being the system Hamiltonian. Substituting $F$ by $H$ in Eq.~\eqref{eq:biased-CRB} we obtain a \emph{quantum} Cram\'er-Rao bound (QCRB) for an estimator with a given bias.

Figure~\ref{fig:figSM2} shows the RMSE of our estimators compared with the limits established by the biased CRB and QCRB, using the respective bias of each estimator. We observe that the RMSE of both Bayesian and NN estimators for the quantum data saturates the CRB for a large region of values of $\Delta$. The lower limit set by the QCRB is smaller and not saturated by our estimators. This is expected, since photon counting does not necessarily represent the most optimal measurement.

\section{Deploying NN predictors on edge devices with TensorFlow Lite}

\begin{figure*}[h!]
\includegraphics[width=0.65\textwidth]{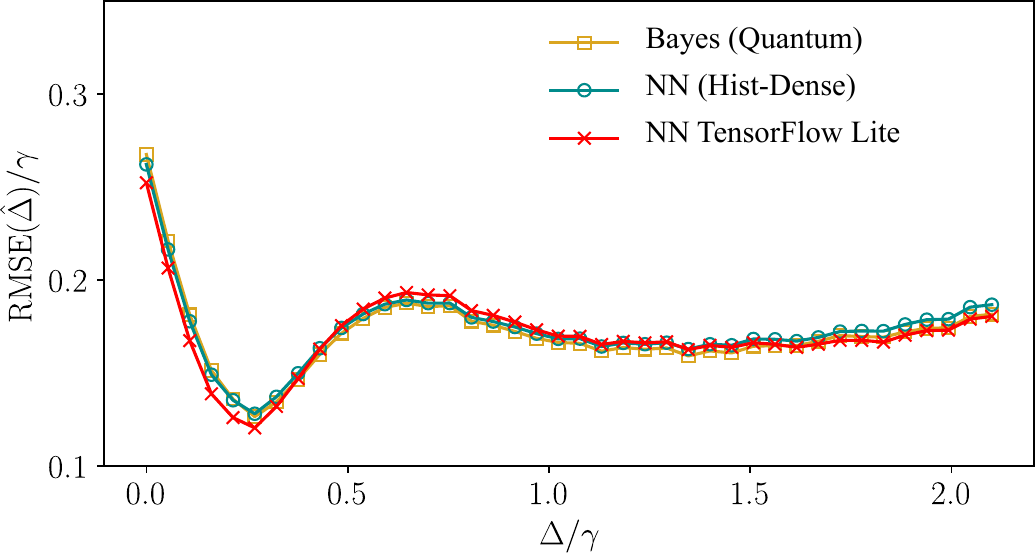}
\caption{\textbf{Inference at the edge.} RMSE of the predictions made by a trained model that has been compressed into a TensorFlow Lite model (red crosses). The RMSE of the Bayesian inference and of the original NN predictor are shown for comparison.}
\label{fig:figSM3}
\end{figure*}

The largest models trained in this work have a size of $\sim 1$MB. While this figure is by no means large, we can further reduce the size by converting these models into compact and efficient data structures using the TensorFlow Lite library. This opens up the opportunity to deploy the models into microcontrollers or other edge devices, making it possible to perform inference on the edge in real time.
Figure~\ref{fig:figSM3} depicts the RMSE of a NN predictor that has been converted into a TensorFlow Lite model weighting only 93kB, showing that the performance of the model remains almost unchanged after the compression, thanks to optimization techniques such as post-training quantization and weights pruning.

\end{document}